
\input phyzzx
\doublespace
\font\zfont = cmss10 scaled \magstep1
\def\ZZ{\hbox{\zfont Z\kern-.4emZ}}


\frontpagetrue

\let\picnaturalsize=N
\def\picsize{1.2in}
\def\picfilename{scipp_tree.eps}

\let\nopictures=Y
\centerline{} \vskip-3pc
\ifx\nopictures Y\else{\ifx\epsfloaded Y\else\input epsf \fi
\let\epsfloaded=Y
{\line{\hbox{
\ifx\picnaturalsize N\epsfxsize \picsize\fi
{\epsfbox{\picfilename}}} \hfill
\vbox{                        
\singlespace
\hbox{SCIPP 95-13}
\hbox{MIT-CTP-2421}
\hbox{SLAC-PUB-95-6776}
\vskip.6in
}
}}}\fi

\rightline{SCIPP 95-13} \vskip-14pt
\rightline{MIT-CTP-2421} \vskip-14pt
\rightline{SLAC-PUB-95-6776}
\vskip10pt


\def\SLAC{\centerline{\it Stanford Linear Accelerator Center}
   \centerline{\it Stanford University}
   \centerline{\it Stanford, CA  94309} }

\def\SCIPP{\centerline {\it Santa Cruz Institute for Particle Physics}
  \centerline{\it University of California, Santa Cruz, CA 95064}}

\def\MIT{\centerline {\it Laboratory for Nuclear Science and Department
         of Physics  }
         \centerline {\it Massachusetts Institute of Technology}
         \centerline {\it Cambridge, MA  02139} }

\vskip-.3in
\title{\seventeenbf Supersymmetry Breaking in the Early Universe}

\centerline{{\fourteencp Michael Dine}
\foot{Work supported in part by the Department of Energy.}}
\singlespace
\SCIPP
\vskip8pt

\centerline{{\fourteencp Lisa Randall}
\foot{NSF Young Investigator Award,
Alfred P.~Sloan
Foundation Fellowship, DOE Outstanding Junior
Investigator Award. Supported in part
by DOE contract DE-AC02-76ER03069 and by NSF grant PHY89-04035.}}
\singlespace
\MIT
\vskip8pt

\centerline{{\fourteencp Scott Thomas}
\foot{Work supported by the Department of Energy under contract
DE-AC03-76SF00515.}}
\singlespace
\SLAC

\vskip.6cm
\vbox{
\centerline{\bf Abstract}

Supersymmetry breaking in the early universe induces scalar soft
potentials with curvature of order the Hubble constant. This has
a dramatic effect on the coherent production of scalar fields along
flat directions.  For the moduli problem it generically gives a
concrete realization of the problem by determining the field value
subsequent to inflation.  However it might suggest a solution if
the minimum of the induced potential coincides with the true minimum.
The induced Hubble scale mass also has important implications for
the Affleck-Dine mechanism of baryogenesis.  This mechanism requires
large squark or slepton expectation values to develop along flat
directions in the early universe.  This is generally not the case if
the induced mass squared is positive, but does occur if it is negative.
The resulting baryon to entropy ratio depends mainly on the dimension
of the nonrenormalizable operator in the superpotential which
stabilizes the flat direction, and the reheat temperature after
inflation.  Unlike the original scenario, it is possible to obtain
an acceptable baryon asymmetry without subsequent entropy releases.
}

\endpage

\parskip 0pt
\parindent 25pt
\overfullrule=0pt
\baselineskip=18pt
\tolerance 3500
\endpage
\pagenumber=1

\def\O{{\cal O}}

\def\({\left (}
\def\){\right )}

\def\mgravitino{m_{3/2}}

\def\O{{\cal O}}

\def\bj{\bar{j}}

\def\K1{K^{(1)}}

\def\intF{\int d^2 \theta}


\REF\polonyi{C. Coughlan, W. Fischler, E. Kolb, S. Raby,
and G. Ross, Phys. Lett. B {\bf 131} (1993) 59; J. Ellis,
D.V. Nanopoulos and M. Quiros, Phys. Lett. B {\bf 174}
(1986) 176.}

\REF\stringmoduli{ R. de Carlos, D.V. Nanopoulos, and
M. Quiros, Phys. Lett. B {\bf 318} (1993) 447;
T. Banks, D. Kaplan, and A. Nelson, Phys. Rev. D {\bf 49}
(1994) 779.}

\REF\bs{T. Banks, M. Berkooz, and P.J. Steinhardt,
preprint RU-94-92, hep-th 9501053.}

\REF\ad{I. Affleck and M. Dine, Nucl. Phys. B {\bf 249} (1985) 361.}

\section{Introduction}

Low energy supersymmetry, if it exists in nature, is likely to
have dramatic consequences for the early universe.
One of the most striking stems from the existence of flat directions
in the scalar potential.  Such directions are a generic
feature of supersymmetric theories, unfamiliar in
conventional field theories.
In string theory, for example,
there are often moduli which label
degenerate classical vacuum states of the string.
These states remain degenerate to all orders in perturbation
theory.
In the minimal supersymmetric standard model (MSSM) there exist,
at the level of renormalizable operators and ignoring
supersymmetry breaking, a large number
of flat directions along which some combination of
squark, slepton, and Higgs fields have expectation values.
In the early universe if the fields parameterizing a flat direction
start displaced from the true minimum, coherent oscillations result
when the Hubble constant becomes smaller than the effective mass.
The energy stored in these oscillations amounts to a condensate of
nonrelativistic particles.
The  production of such condensates should be a
generic feature of supersymmetric theories.
In this paper we discuss the effect of supersymmetry breaking in the
early universe on coherent field production,
with emphasis on the cosmological moduli problem
\refmark{\polonyi-\bs} and Affleck-Dine scenario
for baryogenesis \refmark{\ad}.

Most discussions of the coherent production of scalar fields
assume the potential along
flat directions arises from the same SUSY breaking
responsible for the mass splitting among the standard
model fields in the present universe.
The curvature of the potential would then be set by the
gravitino mass, $V^{\prime \prime} \sim \mgravitino^2$.
If this where the case the field would be highly
overdamped for $H \gg \mgravitino$,
and only begin to oscillate when $H \sim \mgravitino$.
Here we observe that the finite energy density in the early
universe induces a soft potential with curvature
of order the Hubble constant, $V^{\prime \prime} \sim H^2$.
The flat directions are then always parametrically near
critically damped, and efficiently evolve to an instantaneous
minimum of the potential.
For both the moduli problem and AD mechanism, this leads to a
precise way of understanding the ``initial conditions'' for the amplitude
of the fields  when they begin to
oscillate freely at $H \sim \mgravitino$.
In the case of the moduli problem, this suggests a possible solution
if the minimum of the induced
potential coincides with the true minimum.
For the AD mechanism, it gives a much more
complete understanding of the conditions for baryogenesis,
namely a negative mass squared from the finite energy breaking.
This permits an estimate of the asymmetry which systematically
includes the effects of nonrenormalizable terms in the superpotential.
The resulting asymmetry is largely independent of any
assumptions about initial conditions.

\section{Supersymmetry Breaking}

The finite energy density in the early universe 
breaks supersymmetry.
In a thermal phase this is manifest through the
disparate occupation numbers for bosons and fermions.
In an inflationary phase in which a positive vacuum energy dominates,
the inflaton $F$ or $D$ component is necessarily nonzero,
implying supersymmetry breaking.
The same is true in the post-inflationary phase
before reheating, when the inflaton
oscillations dominate, and the time averaged vacuum energy is positive.
In principle, SUSY breaking can be
transmitted to flat directions by both renormalizable and
nonrenormalizable interactions.
However, for large field values all fields which couple through
renormalizable interactions gain a mass
larger than any relevant scale of excitation.
These states then effectively
decouple and do not lift the flat directions.


Nonrenormalizable interactions can have important effects though.
To illustrate this consider a term in the Kahler potential of the
form
$$
\delta K= {1 \over M_p^2}\chi^{\dagger} \chi \phi^{\dagger} \phi
\eqn\deltak
$$
where $\chi$ is a field which dominates the energy density of
the universe, $\phi$ is a canonically normalized flat
direction, and
$M_p = m_p/ \sqrt{8 \pi}$ is the reduced Planck mass.
No symmetry prevents such a term, which can be present directly
at the Planck scale, or be generated by running to a lower scale.
If $\chi$ dominates the
energy density, then
$\rho \simeq \langle \intF \chi^{\dagger} \chi \rangle$.
In a thermal phase the expectation value arises from kinetic
terms over the $\chi$ component thermal distributions.
In an inflaton dominated era it is given by the inflaton
$F$ components and kinetic energy.
The interaction
\deltak~gives an effective mass for $\phi$ of
$
\delta {\cal L} = ( \rho / M_p^2) \phi^{\dagger} \phi
$
(note that a positive contribution in the Kahler potential gives
a negative contribution to $m^2$).
In a flat expanding background, $\rho = 3 H^2 M_p^2$,
so that $m^2 \sim H^2$.
This is a generic result, independent of what specifically dominates
the energy density.
For $H \gsim \mgravitino$, this source for the soft mass is
more important than any hidden sector breaking.

\REF\gravitino{J. Ellis, A. Linde, and D. Nanopoulos, Phys. Lett.
B {\bf 118} (1982) 59;
D. Nanopoulos, K. Olive, and M. Srednicki, Phys. Lett. B
{\bf 127} (1983) 30;
W. Fischler, Phys. Lett. B {\bf 332} (277) 1994.}

In order to be concrete about the evolution along flat directions,
we will assume an inflationary anzatz.
In most models
the correct magnitude of density and temperature fluctuations
in the present universe is obtained
for $H \sim 10^{13-14}$ GeV during inflation.
In order to avoid the gravitino problem the reheat temperature
after inflation can not (conservatively) be larger than about
$10^9$ GeV \refmark{\gravitino}.
This implies that by the era of reheating $H \ll \mgravitino$.
With this restriction the induced potential discussed above is only
important (ignoring an pre-inflationary evolution)
during inflation and in the pre-reheating era dominated by
inflaton oscillations.
We therefore only need to consider the couplings of
the inflaton to the flat directions.

\REF\hamburg{M. Dine, L. Randall, and S. Thomas,
Talk presented at the {\it US-Polish Workshop on Physics from the
Planck Scale to Electroweak Scale}, September 1994; and at the
{\it DESY Theory Workshop on Supersymmetry}, October 1994.}

Since the important couplings between the inflaton and
flat directions arise from Planck scale operators, supergravity
interactions should be included. The supergravity
scalar potential is
$$
V = e^{K/M_p^2} \left(
D_i W K^{i \bj} D_{\bj}W^* - {1 \over 3 M_p^2} |W|^2 \right)
+ {1 \over 8}f_{ab}^{-1} D^a D^b
\eqn\supergravity
$$
where $D_i W \equiv W_i + K_iW/M_p^2$,
$W_i \equiv \partial W / \partial \varphi_i$,
$K^{i \bj} \equiv (K_{i \bj})^{-1}$,
$f_{ab}$ is the gauge kinetic function.
$W(\varphi)$ and $K(\varphi^{\dagger},\varphi)$
and the superpotential and Kahler potential,
$D^a \equiv K_{\varphi} T^a \varphi$,
where $\varphi$ includes in general the flat directions,
inflaton(s), and hidden sector.
If the inflaton potential arises from $F$ terms, the term
in parenthesis has positive expectation value and
a nontrivial potential along flat directions arises.
Even if $D$ terms dominate the inflaton potential, with
nontrivial Kahler potential couplings (such as \deltak)
a potential results.
The general form for the induced potential
from \supergravity~along an exact
flat direction is of the form
$$
V(\phi) = H^2 M_p^2 ~f (\phi / M_p)
\eqn\genericpot
$$
where $f$ is some function.
Notice that the curvature is set by the Hubble constant,
$V^{\prime \prime} \sim H^2$, and the scale for
variations in the potential is $M_p$.
The general lesson is that in the early universe,
when $H \gg \mgravitino$, the characteristic scale for soft
parameters is of order the Hubble constant \refmark{\hamburg}.

\REF\drt{M. Dine, L. Randall, and S. Thomas, to appear.}

In the rest of this letter we describe
some of the consequences of this observation for the moduli
problem and AD mechanism of baryogenesis.
In a forthcoming paper, we will present a much more
detailed discussion of these issues, with particular attention
to the computation of the baryon
asymmetry \refmark{\drt}.

\section{Moduli}

\REF\largemass{The fact that during inflation fields generically
have masses of order $H$ has been noted by many
authors.  In the context of the Polonyi problem,
it was noted by
M. Dine, W. Fischler, and D. Nemeschansky,
Phys. Lett. B {\bf 136} (1984) 169.
In another context, this was noted by
O. Bertolami and G. Ross, Phys. Lett.
B {\bf 183} (1987) 163. }

\REF\dvali{G. Dvali, preprint IFUP-TH 09-95, hep-ph 9503259.}

The coherent production of string moduli leads to the string
version \refmark{\stringmoduli,\bs} of the Polonyi
problem \refmark{\polonyi}.
The late decay of such a condensate can 
lead to a number of cosmological problems, including
modification of the light element abundances.
During inflation the moduli evolve in the potential \genericpot~
with $H \sim$ constant.
Since the fields are parametrically close to critically
damped, within a few $e$-foldings they are driven to
a local minimum of the potential (up to
quantum deSitter fluctuations) \refmark{\largemass,\dvali}.
This is in contrast to the usual assumption that ``scalars
are not diluted during inflation.''
However, the form of the potential does not
necessarily coincide with that after inflation, or from hidden sector
SUSY breaking.
In general the minima are separated by $\O(M_p)$.
Once $H \sim \mgravitino$
the moduli then start to oscillate freely
about a true minimum with amplitude of $\O(M_p)$ \refmark{\drt}.
This just gives a concrete realization of the
initial conditions for the moduli problem by specifying the
field for $H \gsim \mgravitino$.

\REF\rt{L. Randall and S. Thomas, preprint MIT-CTP-2331,
SCIPP 94-16, hep-ph 9407248.}

The present discussion suggests a possible solution of the moduli
problem \refmark{\hamburg}.
If the minima coincide at early and late times the moduli
are driven to the true minimum during inflation.
One possibility under which the minima can coincide
occurs if  there are no Kahler potential couplings
between the moduli and either the inflaton or hidden sectors.
If there is no SUSY preserving nonperturbative superpotential generated
on moduli space, then the potential arises from
supergravity interactions coupling the moduli to $F$ components
in the inflaton or hidden sectors.
Every minimum of the moduli Kahler potential then coincides
with a minimum of the potential at both early and late times
\refmark{\rt}.

\REF\symnote{In the context of the Polonyi problem,
M. Dine, W. Fischler, and D. Nemeschansky,
Phys. Lett. B {\bf 136} (1984) 169
pointed out that symmetries could
give minima which coincide at low and high energies.}

A less restrictive circumstance
under which the minima might coincide
is if there is a point of enhanced symmetry
on moduli space \refmark{\symnote}.
The potential is necessarily an extremum about such points
since the moduli transform under some symmetry.
Such enhanced symmetry points are familiar in
string theory.
In many string compactifications, there are
points in the moduli space
where all of the moduli, with the notable exception of the
dilaton, transform non-trivially under some enhanced
symmetry.

\REF\orbifold{ L. Dixon, J. A. Harvey, C. Vafa, and E. Witten,
Nucl. Phys. B {\bf 261} (1985) 678.}

An example of this phenomenon is provided by the $Z$
orbifold \refmark{\orbifold}.  This orbifold is usually described
by taking a product of three two dimensional tori.  In this construction,
the resulting theory has a variety of symmetries including
an $SU(3)$ gauge symmetry
and two
$Z_3$  $R$ symmetries.
All the moduli are charged under some of these symmetries except those
which describe the three
two-dimensional tori.
At special points in the moduli space there are further enhanced symmetries
under which these remaining
moduli, with the exception of the dilaton, are charged.

It might be that the true ground state of string
theory is near such a point of enhanced symmetry.
Alternately, some or all of these symmetries might be broken by
small $\O(\mgravitino)$
vev's of other fields.
The main problem with this idea
is the dilaton.  It is not known if such an enhanced symmetry exists
for this field, and even if it does, it is likely to lie at a point where the
gauge coupling is extremely large.  So if symmetries are the solution
of the moduli problem, the
dilaton must be on a different footing than the other moduli.  For
example, the dilaton
mass might arise from
dynamics which does not  break supersymmetry.
The serious difficulties which such an idea must face have
been discussed in ref. \refmark{\bs}.
The possibility also exists to solve the moduli problem with a late
inflation \refmark{\rt}.
However unless $H \ll \mgravitino$ the minimum may be shifted as for
standard inflation.

\section{Baryogenesis}

\REF\lh{H. Murayama and T. Yanagida, Phys. Lett. B {\bf 322} (1994) 349.}

\REF\rsymnote{This can in fact be guaranteed by a discrete $Z_N$ $R$
symmetry.  Under such a symmetry the superpotential transforms
as $W \rightarrow e^{4 \pi i / N} W$.  If, for example, $\phi$
transforms as $\phi \rightarrow \phi$, then no terms of the form
$\phi^n$ are allowed. It is possible to forbid terms of the form
$\chi \phi^n$ by symmetries as well;
M. Dine and N. Seiberg, Nucl. Phys. B {\bf 160} (1985) 243.}

\REF\otherad{A. Linde, Phys. Lett. B {\bf 160} (1985) 243;
J. Ellis, D. V. Nanopoulos, Phys. Lett. B {\bf 184} (1987) 37;
K. Enqvist, K.. W. Ng, and K. Olive, Phys. Rev. D {\bf 37}
(1988) 2111;
S. Davidson, H. Murayama, and K. Olive, Phys. Lett. B {\bf 328}
(1994) 354.}

In the MSSM, at the level of renormalizable operators, there
are numerous flat directions in the space of scalar fields.
Most of these
involve squarks or sleptons and carry $B$ and/or $L$.
A simple example is the direction \refmark{\lh}
$$
H_u = \left ( \matrix{\phi \cr 0} \right ) ~~~~~
L = \left ( \matrix{0 \cr \phi} \right ) ~~~~~
\eqn\hldirection
$$
where $\phi$ parameterizes the flat direction.
In the original discussion of ref. \refmark{\ad}
it was assumed that directions
such as this were exactly flat in the supersymmetric
limit \refmark{\rsymnote} and that $\phi$ was initially $\O(M_{GUT})$ or
$\O(M_p)$.
For $H \lsim \mgravitino$ the field would begin to oscillate
about the true minimum at $\phi =0$.
In addition to the $B$ and $L$ conserving terms, the
soft potential was assumed to contain $B$ and/or $L$ violating
dimension four terms suppressed by
$\mgravitino^2/M_p^2$.  As a result, the coherently oscillating field
develops a large baryon number.
The subsequent decay of the condensate
then gives a substantial (even enormous)
baryon asymmetry \refmark{\ad,\lh,\otherad}.

\REF\nonrenorm{J. Ellis, K. Enqvist, D. V. Nanopoulos, and
K. Olive, Phys. Lett. B {\bf 191} (1987) 343;
K. W. Wong, Nucl. Phys. B {\bf 321} (1989) 528.}

With the inclusion of nonrenormalizable terms in
the superpotential, and the induced soft potential,
the scenario for AD baryogenesis is very different.
Nonrenormalizable terms in the superpotential,
if present, will lift flat directions even in the supersymmetric limit.
These can take the form
$$
\delta W = {\lambda \over n M^{n-3}}\phi^n
\eqn\lifting
$$
where $M$ is some large mass scale such as the GUT or Planck scale.
For the $LH_u$ example given above the lowest order term of this form,
assuming $R$ parity, is ${\lambda \over M} (LH_u)^2$.
The power law growth in the potential from these terms
limits the fields
to be parametrically less than $M_p$ (even for $M \sim M_p$).
In addition, $A$ terms, proportional to $\phi^n$,
can result from cross terms in \supergravity~and higher
order terms in the Kahler potential.
In light of our discussion of early universe SUSY breaking,
the scalar potential for $H \gg \mgravitino$ then has the form
\refmark{\hamburg}
$$
V(\phi) \simeq cH^2 \vert \phi \vert^2
+{a\lambda H \phi^n \over n M^{n-3}}
+|\lambda|^2{\vert \phi \vert^{2n-2} \over M^{2n-6}}~~.
\eqn\vphi
$$
where
$c$ and $a$ are constants of $\O(1)$.
The $A$ term has the important effect of violating $B$ or $L$ and
has a definite $CP$ violating phase relative to $\phi$.

\REF\desitter{T. Bunch and P. Davies, Proc. R. Soc. A {\bf 360} (1978)
117;
A. Linde, Phys. Lett. B {\bf 116} (1982) 335;
A. Starobinsky, Phys. Lett. B {\bf 117} (1982) 175;
A. Linde, Phys. Lett. B {\bf 131} (1983) 330.}

\REF\positivemass{The Kahler potential might be tuned to give
$m^2 \ll H^2$ but positive. In that case
large fluctuations can develop with a correlation
length of order
$ H^{-1} e^{3H^2 / 2m^2}$ \refmark{\otherad,\desitter}.
To be coherent over a scale large compared to the current
horizon size requires
$(H /m)^2 \gsim 40$.
However if $m^2 \sim H^2$ after inflation, the field is driven toward
the origin as a power law in time.
So an entirely
positive $m^2$ scenario must fine tune the mass even after
inflation.}

With minimal Kahler potential, the coefficient $c$ arising from
\supergravity~is positive ($c=3$ during inflation for $F$
type inflaton breaking).
The flat direction is then driven exponentially
quickly to the origin during inflation.
Quantum deSitter fluctuations give
$\langle \delta \phi^2 \rangle \sim H^2$, but with a
correlation
length of $\O(H^{-1})$ \refmark{\desitter}.
Any resulting baryon number then averages
to zero over the present universe \refmark{\positivemass}.
In addition the relative
magnitude of the $B$ violating term in \vphi~is small for $H \ll M$.

A non-negligible baryon number can result if the $B$ violating
term in \vphi~has the same magnitude as the $B$ conserving
terms.
This will occur if $c < 0$.
This is perfectly possible for suitable choices of the
Kahler potential; no fine-tuning is required.
In this case the minimum of the potential, ignoring for the
moment the $A$ term, is given by
$$
|\phi_0|=
\({ \sqrt{-c} H M^{n-3} \over (n-1) \lambda} \)^{1 \over n-2}.
\eqn\phizero
$$
Inclusion of the contribution of the $A$ term does not substantially
change the magnitude of the minimum,
but does give $n$ discrete minima for the phase of $\phi$.
During inflation 
if $|c|$ is not too small,
the system quickly settles into one of the minima.
The observable universe is then left with a single value
of the intial phase of $\phi$.
After inflation, $H$ changes with time as in a matter dominated
universe and $\phi_0$ decreases.
A straightforward analysis of the equations of motion
in this era indicates
that for $n \geq 4$, the field oscillates about a point
slightly larger than $\phi_0(t)$
where $V^{\prime \prime}(\phi) \sim H^2$.
Thus when $H \sim m_{3/2}$,
$\langle \phi \rangle \sim \phi_0$.
At this time the soft potential from hidden sector SUSY breaking
becomes important.
The $A$ term from this source
is comparable in magnitude to the other terms in the potential
(as may be seen by simply plugging $\phi_0$ into eq. \vphi)
and in general has a different phase
than any arising from coupling to the inflaton.
The $B$ or $L$ violation is therefore maximal during the epoch at which
the field begins to oscillate freely, thereby imparting a substantial
asymmetry to the condensate.
The resulting baryon number per condensate
particle is near maximal, $n_b / n_{\phi} \sim \O(10^{-1})$
(if the relative phases are $\O(1)$).
Notice that this is {\it independent} of $\lambda / M$.
Once $H \ll \mgravitino$ the field value decreases and the
relative importance of the $A$ term is reduced.
The baryon number imparted to the condensate is
therefore conserved in this epoch.
This scenario has been checked by numerical integration of the
equations of motion \refmark{\drt}.

The baryon to entropy ratio depends on
the total density in the condensate, and
the inflaton reheat temperature , $T_R$.
The flat direction
$\phi$ begins to oscillate freely when the coherent oscillations
of the inflaton still dominate the total energy density,
$\rho_I$.
Since $\rho_{\phi} \sim \mgravitino^2 \phi_0^2$, the
fractional energy in the condensate is
$$
{\rho_{\phi} \over \rho_I} \approx
\left ( {m_{3/2} M^{n-3} \over \lambda M_p^{n-2}}
\right )^{2/(n-2)}.
\eqn\energyratio
$$
Notice that
$\rho_{\phi}$ is larger for smaller
$(\lambda / M^{n-3})$.
After the inflaton decays the baryon to entropy ratio is then
$$
{n_b \over {s}} \approx {n_b \over n_{\phi}}{T_R \over
m_{\phi}}{\rho_{\phi} \over \rho_I}.
\eqn\bns
$$
This estimate is insensitive to the details of the
decay of the AD flat direction, as long as it has nonzero $B-L$.
The baryon to entropy ratio depends mainly on $T_R$
and the order at which the flat
direction is lifted.
For $T_R$ just below the gravitino bound and $M \sim M_p$,
$n_b/s$ is too large for $n \geq 6$.
However for the $LH_u$ direction with $n=4$,
after sphaleron processing of the resulting lepton number,
$n_b /s \sim 10^{-10}(T_R / 10^9 {\rm GeV})
( M/ \lambda M_p)$.
This is a quite reasonable range.
At low energies the operator ${\lambda \over M}(LH_u)^2$ which lifts this
flat direction gives rise to neutrino masses.
In this scenario $n_b /s $ can therefore be related to the {\it lightest}
neutrino mass since the field moves out furthest along the eigenvector
of $L_i L_j$ corresponding to the smallest eigenvalue of the
neutrino mass matrix,
 $n_b /s \sim 10^{-10}(T_R / 10^9 {\rm GeV})
( 10^{-5} {\rm eV} / m_{\nu})$.
The total baryon density in the condensate grows rapidly
with $n$; only the $LH_u$ direction gives a reasonable result without
additional entropy releases after inflaton decay.




\section{Conclusions}

In summary
the large supersymmetry breaking in the early universe
gives a precise realization of
the ``initial conditions" (when $H \sim \mgravitino$)
along flat directions.
It seems quite difficult to
solve the moduli problem unless there are symmetries which
ensure that the high energy and low energy potentials
possess the same minimum.  We have seen that
(much to the suprise of some of the authors) the AD mechanism
is quite robust.  Provided that the curvature of the
induced $\phi$ potential
at the origin is negative for $H \gg \mgravitino$,
a desirable value for $n_b/s$ results for the $LH_u$ direction
when account is taken of higher dimension operators.  More
detail about the evolution of the fields,
other standard model flat directions,
the possible sources of
supersymmetry breaking, and the decay of the condensate
will be presented in ref. \refmark{\drt}.

\refout
\bye